# Focused ion beam preparation of atom probe specimens containing a single crystallographically well-defined grain boundary

Fabián Pérez-Willard[*,1], Daniel Wolde-Giorgis[2], Talaát Al-Kassab[2], Gabriel A. López[3], Eric J. Mittemeijer[3], Reiner Kirchheim[2], and Dagmar Gerthsen[1]

[1] *Laboratorium für Elektronenmikroskopie, Universität Karlsruhe (TH), D-76131 Karlsruhe, Germany*

[2] *Institut für Materialphysik, Universität Göttingen, D-37077 Göttingen, Germany*

[3] *Max Planck Institute for Metals Research and Institute for Physical Metallurgy, University of Stuttgart, D-70569 Stuttgart, Germany*

**Needle-shaped atom probe specimens containing a single grain boundary were produced using the focused ion beam (FIB) of a dual-beam FIB/SEM (scanning electron microscope) system. The presented specimen preparation approach allows the unprecedented study of a grain boundary which is well characterised in its crystallographic orientation by means of the field ion microscope (FIM) and the tomographic atom probe (TAP). The analysis of such specimens allows in particular the determination of solute excess atoms at this specific grain boundary and hence the investigation of the segregation behaviour. The crucial preparation steps are discussed in detail in the present study for the $\Sigma$ 19a {331} <110> grain boundary of a 40 at.ppm-Bi doped Cu bi-crystal. Transmission electron microscope (TEM) images and TAP analyses of the atom probe tips demonstrate unambiguously the presence of the selectively prepared grain boundary in the apex region of some of the specimens.**

[*] *Correspondence should be addressed to:*

*Fabián Pérez-Willard, Engesserstr. 7, D-76131 Karlsruhe, Germany*
*Tel.: +49 (0)721 608 8681, Fax: +49 (0)721 608 3721*
*e-mail: fabian.perez@physik.uni-karlsruhe.de*

# 1. Introduction

The atom probe field ion microscope (APFIM) is a powerful analytical tool that allows not only to see, but also to identify the chemical nature of individual atoms of a sharp conducting tip [1]. In combination with a position sensitive detector, in the so-called tomographic atom probe (TAP), a three dimensional reconstruction atom-by-atom of the sample can be achieved [2]. Imaging and analysis with the APFIM require extremely high electric fields of 20-50 V/nm at the tip surface in order to attain field ionisation (FIM mode) or field evaporation (atom probe mode). By applying a high voltage to the tip these high electric fields can be reached only at a tip apex of hemispherical shape provided its radius of curvature is below 100 nm. The preparation of suitable tips remains the main challenge for any kind of scientific study with the APFIM.

The most simple and convenient method to prepare FIM/TAP tips is still to apply conventional electro-polishing techniques to a bulk rod sample typically having a diameter of 200 µm and a length of 15 mm [1,2]. This technique has been used for many decades for a successful preparation of FIM tips of bulk materials. However, the method may fail when applied to bulk nano-crystalline materials since preferential etching along grain boundaries becomes predominant and prevents the formation of a suitable hemispherical apex. This is a distinct disadvantage since the development and application of special functional modern materials such as thin film layers and nano-crystalline materials have become recently of considerable interest for scientific investigations. Hence, different alternative preparation techniques have been proposed including the use of pre-shaped substrate tips [3] and the use of electron-beam lithography methods [4] or FIB to fabricate suitable tips from ball-milled powder specimens and planar grown thin film layers [5,6,7].

Recently, an alternative FIB-based lift-out method has been proposed by Miller *et al.* [8], which allows a site-specific preparation of tips from bulk samples. Miller *et al.* demonstrated that tips can be prepared from thin ribbons, sheets or powders. However, a particular challenge presents the preparation of tips containing a grain boundary to study e.g. segregation phenomena in metallic alloys. Using an alternative approach Colijn *et al.* [9] were able to prepare a tip which contained a grain boundary from an aluminium alloy. Their procedure started with a wedge-shaped thin slice of the material. The tip remained in the original slice without the need of final lift out because a local electrode atom probe was used. However, a standard APFIM requires lift out of the tip which has to be firmly attached to a blunted tungsten wire.

Our work was motivated by the need to prepare tips for a standard APFIM from special well-oriented grain boundaries to study quantitatively segregation phenomena in metallic alloys. Since the amount of excess solute atoms depends strongly on the nature of the grain boundary (i.e. the orientation relationship of both grains and the grain boundary plane) a quantitative and selective analysis of the segregation behaviour of solute atoms at special grain boundaries is required for a satisfactory theoretical description. We present here a dedicated route, which allows the preparation of FIM/TAP tips containing a single well-defined grain boundary from a bulk bi-crystalline sample. This task is particularly challenging because the bi-crystalline sample contains only one grain boundary as opposed to numerous grain boundaries in a polycrystalline sample. For the analysis of the grain boundary in a conventional FIM/TAP setup, the tip has to be lifted out of the original bulk sample and attached to a blunted tungsten tip.

**2. Experimental methods**

The preparation steps are presented for the Σ 19a {331} <110> grain boundary of a 40 at.ppm-Bi doped Cu bi-crystal, which was grown by the Bridgman technique and doped with Bi from a dilute vapour phase. The sample was then homogenised at 950 °C for 10 d in an Ar atmosphere. The bulk composition of 40 at.ppm Bi was measured by means of inductively coupled plasma optical emission spectrometry (ICP-OES). Finally, a heat treatment at 850 °C for 12 h was performed in order to achieve the equilibrium segregation of Bi at the grain boundary.

Before mounting the bi-crystal in the FIB/SEM system its surface was carefully ground and polished, and etched in 65% $HNO_3$ for approximately 2-3s. After etching the grain boundary is clearly visible under the optical microscope and in the SEM.

Our sample preparation requires a tungsten needle as a support. The W supports were prepared by standard electro-polishing. A 2 molar NaOH solution in $H_2O$ was used as an electrolyte and an alternating voltage in the range of 5-10 V was applied between the W needle and the second electrode.

The FIB work was performed by means of a dual-beam FIB/SEM system 1540XB from Carl Zeiss SMT. The focused ion beam consists of $Ga^+$ ions accelerated by a voltage of 30 kV. In addition, the FIB/SEM system is equipped with an *in-situ* gas injection system and an *in-situ* micromanipulator.

TEM analyses of the FIM tips were performed using a specially designed specimen holder in a 120kV Philips EM-400 electron microscope.

The TAP characterisation of the tips was done in a FIM/TAP instrument equipped with a position sensitive detector from Cameca, which allows the reconstruction of a region of the tip with a lateral spatial resolution of 0.5 nm.

## 3. Accessing the grain boundary

The bicrystal surface shows tracks which reveal the position of the grain boundary and also the relative crystallographic orientation of the two single crystals (see Fig. 1(a)). The tracks on each side of the grain boundary form an angle of 153.5° as expected for the Σ 19a {331} <110> grain boundary. They are a result of the anisotropic etching of Cu along preferential crystallographic planes in a $HNO_3$:$H_2O$ solution.

After the chemical etching the two single crystals form a step at the grain boundary with a height that can reach one to a few microns (see Fig. 1(b)). In a region around the grain boundary this step was levelled to create a flat platform by performing successive striping FIB cuts with currents of $I_{Ga}$ = 500 and 200 pA. The result is shown in Fig. 1(c). Some pores can be observed at the grain boundary due to local disturbances during the growth of the bicrystal. For convenience, only regions in which the grain boundary follows almost a straight line were prepared in this way.

A platinum sacrificial layer is deposited on the chosen region of the grain boundary to reduce the rounding of the top of the specimen and minimise beam damage during the FIB milling. This deposition process consists of an electron-beam induced deposition (EBID) of approximately 250 nm height and 1-2 µm width and a second stage of protective platinum layer by a much faster ion beam induced deposition (IBID) to reach a total height of approximately 1 µm. Trenches are then milled parallel to the grain boundary plane and on both sides of the platinum deposit to create a 1-2 µm thick lamella (see Fig. 2(a)) in analogy to the initial steps of the standard TEM lift-out sample preparation scheme [11,12]. In our case, the trenches were milled using FIB currents of $I_{Ga}$ = 10 nA, 2 nA and 500 pA with decreasing current as the platinum deposit was approached. For the 500 pA cuts the sample tilt was increased by 1° in order to improve the verticality of the side walls of the lamella. The dimensions of the lamella shown in Fig. 2(a) are 20 µm x 2 µm x 10 µm. In a next step the sample is tilted to an angle of 51° between the surface normal and the ion beam. The bottom and left side of the lamella is then cut free ( $I_{Ga}$ = 200 pA, c.f. Fig. 2(b)).

With the aid of an *in-situ* micromanipulator, small pillars containing the grain boundary can be lifted out of the sample in a procedure similar to the one described in detail in Ref. [8]. In brief: the tip of the micromanipulator is attached by IBID of tungsten to an edge of the lamella, a long pillar with a cross section area of roughly 2 µm x 2 µm is cut free from the lamella (see Fig. 2(b)) and lifted-out by lowering the microscope stage (see Fig. 2(c)).

Finally, the pillars are attached to a tungsten support, consisting of a W tip which has been previously blunted (in an earlier FIB session). This is illustrated in Figs. 2(d) and 2(e). In Fig. 2(d) the pillar and the W post have been carefully aligned by using the stage for a rough and the micromanipulator for the fine positioning. The pillar is attached to the W support with an IBID of either W or Pt ($I_{Ga}$ = 10-50 pA) on two sides of the pillar and then cut free from the micromanipulator tip (c.f. Fig. 2(e)). As described in the next section, in a subsequent step the pillar is shaped to form a sharp needle. To ensure that the specimen remains firmly attached to the W support after this next step, it is important that the IBID completely fills the gap between the pillar base and the W support in the region of interest. Figure 2(f) shows a top view of a pillar containing the grain boundary attached to its W support.

## 4. Shaping of the FIM/TAP specimen

In this section, the FIB shaping of the pillar in the form of a sharp needle is described. The main difficulty here is to keep track of the location of the grain boundary during the different milling steps which are necessary to define the FIM/TAP tip. A typical FIB-produced FIM/TAP tip is shown in Fig. 3(a). Ideally the grain boundary should run from the base to the apex of the tip. This is by far non-trivial, because of the small radius of curvature below 100 nm of the finished FIM/TAP tip. In addition, the two differently oriented copper grains exhibit a poor contrast in the SEM.

In order to visualise the grain boundary one side of the pillar is polished at low FIB currents ($I_{Ga}$ <= 50 pA) (further explanation is given in section 6, compare with Fig. 4(c)). The image must be carefully analysed in order to discriminate the grain boundary from curtain effects produced by milling through the sacrificial platinum deposit. Once the grain boundary has been localised within the pillar in the SEM image (and FIB image) line cuts with the ion beam slowly approaching the grain boundary from the left, from the right and from the back, respectively are performed at low currents ($I_{Ga} \leq 50$ pA) to define the final shape of the FIM/TAP specimen (cf. section 6 and Figs. 4(c) and 4(d)). A radius of curvature of the apex of 20-40 nm and shaft angles (half-opening angles) in the range of 2-8° can be achieved routinely. Because of the dependence of the sputter yield on the angle of incidence of the ion beam and the increased sputter yield at edges, the top-most region of the finished specimens automatically have a cylindrical symmetry (compare with Figs. 5(b)-(c)). A typical FIB-produced FIM/TAP tip is shown in Fig. 3(a).

During the sharpening of the specimen special care has to be taken to avoid producing accidental needles near the specimen. This important issue will be discussed in detail in the next section.

**5. Simulations**

An issue of concern are the sharp edges of the W support and the accidental needles near the specimen characteristic of the above described FIB fabrication process. They may cause artefacts in the FIM images and positioning errors in the TAP measurements. This is straightforward because the sample itself represents the major optics in the FIM/TAP microscope. Therefore, numerical simulations have been carried out in order to elucidate the influence of the given specimen geometry [13]. The qualitative simulations of the electric field distribution were performed using the electron and ion optics simulation program SIMION v7.0 [10]. Given a conductive specimen at the voltage $U$, the algorithm calculates the surrounding electrostatic potential landscape by solving the Laplace equations. The arrangement of the tip apex, shaft and fixing post was modelled using an array of 50 millions points. Two questions of special relevance arise: a) Will the edges and accidental tips ionise the imaging gas in the FIM mode or even cause unintentional field-induced evaporation of material? b) Will they have an influence on the trajectories of the ionised atoms?.

Fig. 3(a) and 3(b) show a typical tip after the FIB fabrication process and the assumed geometry in our simulation, respectively. The edges of the W post are modelled as 90° edges. The accidental needles are simulated as a ring enclosing the main tip. Thus, the whole object has a cylindrical geometry, which minimises computation time. In Figs. 3(c), (d) and (e) the equipotential lines (in red) surrounding the FIM/TAP tip have been plotted for the cases of a stand-alone tip, a tip on a post, and a tip on a post with the enclosing ring, respectively. In all three cases the tip was assumed to have a radius of curvature of 30 nm, a length of 19.2 µm and a shaft angle of 8.5°. The tip was assumed to be at a potential of 10 kV, which is a realistic value for a FIM or TAP measurement. In the figures the potential difference of neighbouring equipotential lines equals 1 kV.

An electric field of approx. 30 V/nm is necessary to ionise the Cu atoms of the specimen in the TAP mode, or the atoms of the imaging gas (typically neon) in the FIM mode [1]. As a consequence (see Fig. 3(f)), in the simulation ionisation is only possible in a small region around the tip apex for both, a stand-alone tip (region limited by the red line), and a tip plus post plus ring (region limited by the blue line). In the latter case the ionisation region has shrunk significantly, thus limiting the region of the sample which can be imaged and analyzed in the FIM/TAP experiment. Furthermore, in the presence of a W post and accidental tips, the equipotential lines separate from the specimen shaft at a smaller distance from the apex (see Fig. 3(c)-(e)), and therefore the effective shaft angle of the sample is increased. In the TAP measurement a bigger shaft angle translates in a voltage increase at a higher rate, in order to field evaporate successive layers of material.

Another issue of concern is the influence of the surroundings of the tip on the trajectories of the ionised particles. The trajectories of particles starting from different regions of the specimen were calculated for our three models (see e.g. Fig. 3(d)). Due to their inertial mass the ions follow only roughly the gradient of the electric potential. The potential landscape in the immediate neighbourhood of the tip apex is identical for the three models considered. Thus, while trajectories of ions starting from the shaft region of the tip can deviate considerably in the three cases, the deviations in trajectory are negligible for those ions starting from the real ionisation region.

Summarising the results of our simulations: Obviously, in the presence of edges and accidental tips the electrostatic potential landscape surrounding the specimen tip changes. Fortunately, these changes are negligible in the immediate neighbourhood of the tip apex, i.e. in the region where ionisation takes place. Therefore, they do not lead to positioning errors in the FIM/TAP experiment. On the other hand, sharp edges and accidental tips modify the potential landscape around the tip shaft in a way, which qualitatively corresponds to an increased effective shaft angle of the specimen. Thus, the ionisation region at the tip apex shrinks for a given value of $U$ and with it also the size of the analysable sample region in the FIM/TAP experiment. With this in mind the FIB preparation process was optimised as described in the next section to avoid edges and accidental needles near the specimen.

**6. Optimised FIM/TAP specimens**

In this section, two different approaches which have been pursued in order to optimise the FIM/TAP specimen geometry are presented. In both cases, the main idea is to try to mimic the ideal case of a stand alone tip. Sharp edges are completely avoided and the accidental tips inherent to the FIB fabrication process are located at a larger distance from the specimen apex.

In the first approach the W support is integrated into the specimen by shaping the former with annular FIB cuts. The procedure is illustrated in Figure 4 for a selected specimen (see Fig. 4(a)). The specimen is tilted towards the FIB column and then annular cuts centred at the assumed grain boundary location are performed at ion currents in the range of 500 pA to 2 nA. The annular cuts aim to eliminate the protruding parts of the W support under the pillar (see Fig. 4(b)). After the shaping of the support the specimen fabrication process follows the steps described already in section 4. Figure 4(c) shows the specimen after the polishing at low FIB currents. The grain boundary is clearly visible in the SEM image taken with the in-lens SE detector. Figs. 4(d) and (e) show the sample during the final stages of preparation and a TEM image of the apex region of the finished sample (prior to FIM/TAP analysis), respectively. The grain boundary is oriented almost parallel to the specimen axis all the way to its top.

In an alternative approach illustrated in Figure 5 very slender W supporting tips have been chosen. The cross section of the W support is in this case smaller than that of the pillar. Consequently, a shaping of the W support is no longer necessary. A minor handicap of this approach is that some practice is needed for the precise alignment of pillar and support after the lift-out step (compare Figs. 2 (d)-(e)).

As discussed in the next section, the tips fabricated by either of the two approaches are best suited to perform FIM and TAP studies of selected crystallographically well-defined grain boundaries.

**7. TAP measurements and discussion**

FIM/TAP measurements were performed for a number of FIB prepared samples with non-optimised and optimised geometries [13]. Figure 6 (left-hand side) shows the three-dimensional TAP reconstruction map of one of the non-optimised specimens (i. e. a sample, as in Fig. 3(a), in which the tungsten support had not been integrated to the specimen by FIB shaping). The region depicted is roughly 25 x 7 x 7 $nm^3$. Each of the blue dots represents an individual Bi atom while, for clarity, the Cu-atoms are not shown. The presence of the grain boundary in the sample is revealed by the distribution of the Bi atoms and most clearly seen in the coloured planar concentration profile on the right-hand side of Figure 6. The Bi enriched region has a surface of approximately 148 $nm^2$ and a width of 2-3 nm. This region is assigned to the grain boundary. The number of Bi counts —taking the detection efficiency into account— yield a concentration of solute excess Bi atoms of $(3.2 \pm 0.5)$ atoms/$nm^2$ in the grain boundary [13]. Similar values have been obtained in earlier energy dispersive X-ray spectroscopy TEM studies of the $\Sigma$ 19a {331} <110> grain boundary of similarly Bi-doped Cu bicrystals [14].

For samples where the grain boundary was clearly missed, the Bi signal was below the detection level of 200 at.ppm for the TAP analyses, and only Cu ions were detected. None of the samples studied showed a significant concentration level of Ga atoms, which are expected to be implanted to a depth of some 10 nanometers or less into the specimen during the FIB fabrication process. A possible explanation for this is based on technical reasons, since the TAP setup used in this study only allows data acquisition at voltages above 2.3 kV. Below this voltage the topmost atomic layers are already removed, which results in a clean and uncontaminated specimen before the TAP analysis is started.

After optimisation of the tip geometry it was found empirically that the high voltage applied to the tip needed to produce field ionisation or field evaporation is roughly a factor of two smaller than for the non-optimised specimens. As a consequence, electric forces acting on the tip, which can cause the detachment of the tip from its support, are smaller. The life-time of the optimised tips is increased considerably and switching between FIM and TAP mode during the analysis of the samples is less problematic. Most importantly, for the optimised samples the size of the analysable region is bigger —which is consistent with the simulations results (see section 5)— and thus, the probability of hitting the grain boundary increases.

## 8. Summary

Atom probe specimens were prepared from a bulk bi-crystalline Cu sample containing a single $\Sigma$ 19a {331} <110> grain boundary of a 40 at.ppm Bi-doped Cu bi-crystal by means of a focused ion beam lift-out technique. Simulations were performed in order to study qualitatively the influence of accidental tips inherent to the FIB fabrication process in the FIM/TAP experiment. The simulation results motivate our efforts to optimise the geometry of the FIB-fabricated specimens. The idea underlying the optimised process is to mimic the case of a stand alone tip. TEM and TAP analyses of our specimens show unambiguously the presence of the grain boundary in the sample apex. A concentration of solute excess Bi atoms of $(3.2 \pm 0.5)$ atoms/nm$^2$ for the $\Sigma$ 19a {331} <110> grain boundary was determined. The FIB-prepared specimens allow the unprecedented quantitative TAP study of those phenomena related to grain boundaries for which a precise knowledge of the grain boundary solute excess is required. The detailed experimental analysis of the grain boundary area, made possible by the proposed specimen preparation technique for the APFIM, will for the first time allow validation of rigorous thermodynamical treatments developed for grain-boundary segregation.

**Acknowledgements**

The financial support by the Deutsche Forschungsgemeinschaft (DFG) through SFB 602-B1, KI 230/28-1 and within the Center for Functional Nanostructures (CFN) is acknowledged.

# References

[1]     Müller, E.W., Tsong, T.T., 1969. Field Ion Microscopy - Principles and Applications, American Elsevier Publ Comp, New York.

[2]     Miller, M.K., Smith, G.D.W., 1989. Atom Probe Microanalysis - Principles and Applications to Materials Problems, Materials Research Society, Pittsburgh.

[3]     Al-Kassab, T., Macht, M.-P., Wollenberger, H., 1995. FIM/TAP analysis of Cu-Pd multilayers. Appl. Surf. Sci. 87/88, 329--336.

[4]     Hasegawa, N., Hono, K., Okano, R., Fujimori, H., Sakurai, T., 1993. A method for preparing atom probe specimens for nanoscale compositional analysis of metallic thin films. Appl. Surf. Sci. 67, 407--412.

[5]     Ohsaki, S., Hono, K., Hidaka, H., Takaki, S., 2004. Focused ion beam fabrication of field-ion microscope specimens from mechanically milled pearlitic steel powder. J. Electron Microscopy 53, 523--525.

[6]     Larson, D.J., Foord, D.T., Petford-Long, A.K., Anthony, T.C., Rozdilsky, I.M., Cerezo, A., Smith, G.W.D., 1998. Focused-ion-beam milling for field-ion specimen preparation: preliminary investigations. Ultramicroscopy 75, 147--159.

[7]     Larson, D.J., Foord, D.T., Petford-Long, A.K., Liew, H., Blamire, M.G., Cerezo, A., Smith, G.W.D., 1999. Field-ion specimen preparation using focused ion-beam milling. Ultramicroscopy 79, 287--293.

[8]     Miller, M.K., Russell, K.F., Thompson, G.B., 2005. Strategies for fabricationg atom probe specimens with a dual beam FIB. Ultramicroscopy 102, 287--298.

[9]     Colijn, H.O., Kelly, T.F., Ulfig, R.M., Buchheit, R.G., 2004. Site-specific FIB preparation of atom probe samples. Microsc. Microanal. 10 (Suppl. 2), 1150--1151.

[10]    For more information, please refer to the SIMION homepage: *www.simion.com*

[11]    Overwijk, M.H.F., van der Heuvel, F.C., Bulle-Lieuwma, C.W.T., 1993. Novel scheme for the preparation of transmission electron microscopy specimens with a focused ion beam. J. Vac. Sci. Technol. B11, 2021--2024.

[12]    Giannuzzi, L.A., Stevie, F.A., 1999. A review of focused ion beam milling techniques for TEM specimen preparation. Micron 30, 197--204.

[13]    Wolde-Giorgis, D., 2005. Grain boundary segregation in silver-nickel and copper-bismuth alloys. PhD thesis, Göttingen.

[14]    Sigle, W., Shang, L.-S., Gust, W., 2002. On the correlation between grain-boundary segregation, faceting and embrittlement in Bi-doped Cu. Phil. Mag. A, 1595--1608.

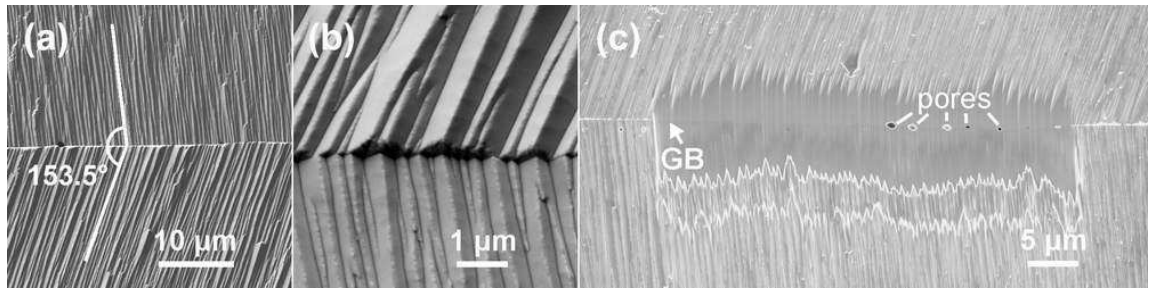

**Figure 1** Accessing the grain boundary (I). (a) Top view of the bicrystal. The tracks on each side of the grain boundary caused by preferential etching form an angle of 153.5° as expected for the sigma 19a grain boundary. (b) 45° tilt view of the grain boundary. The two single crystals form a step at the grain boundary. (c) Top view of a grain boundary region that has been levelled with FIB striping cuts.

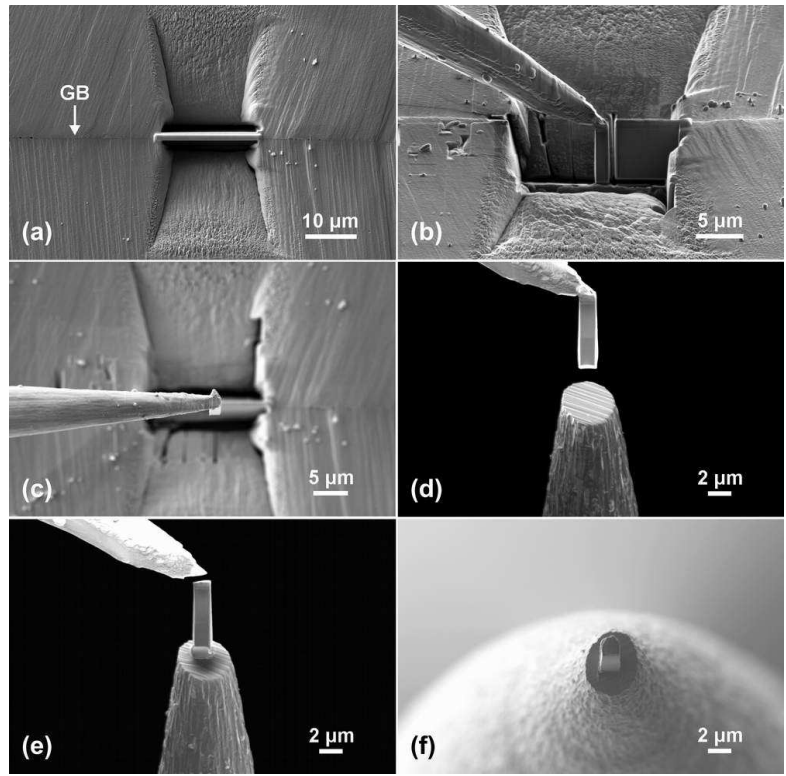

**Figure 2** Accessing the grain boundary (II). (a) Fabrication of a thick lamella containing the grain boundary. (b)-(c) A pillar has been cut free from the lamella and is lifted-out *in-situ* with a micromanipulator. (d)-(e) Attachment of the lifted pillar to a tungsten support, consisting of a sharp W tip that has been previously blunted by FIB. (f) View of the pillar and its W support. The panels (b), (d) and (e) correspond to FIB-induced SE images.

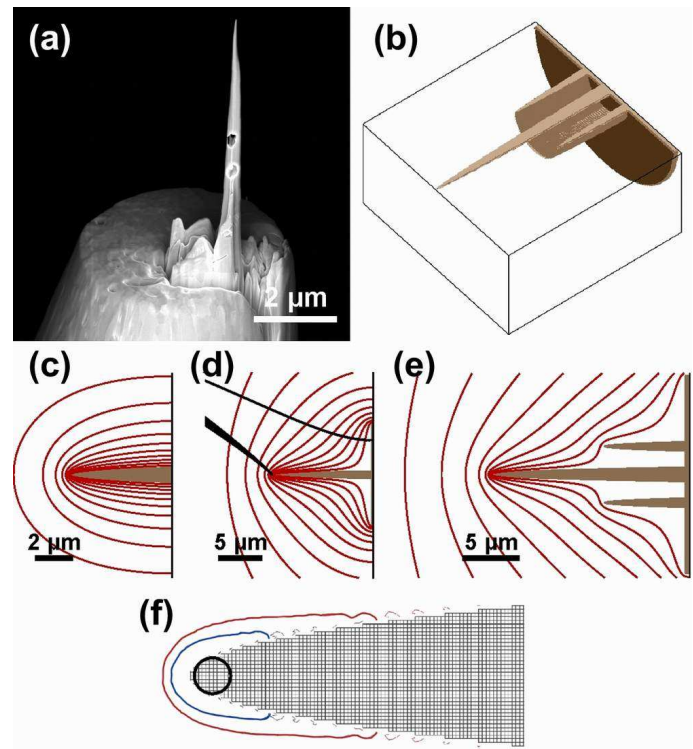

**Figure 3** (a) Image of a typical FIB-shaped FIM/TAP tip. (b) Assumed geometry in the simulation of the electrical potential. The accidental needles are simulated as a ring enclosing the main tip. The whole object has a cylindrical geometry. (c), (d), (e) The three models considered in this study: a stand-alone tip, a tip on a post, and a tip on a post with the enclosing ring, respectively. The red lines show the equipotential lines surrounding the FIM/TAP tip. In panel (d) the trajectories of particles starting from different regions of the specimen are also shown (black lines). (f) Regions in which ionisation is possible for a stand-alone tip (region limited by the red line), and a tip plus post plus ring (blue line).

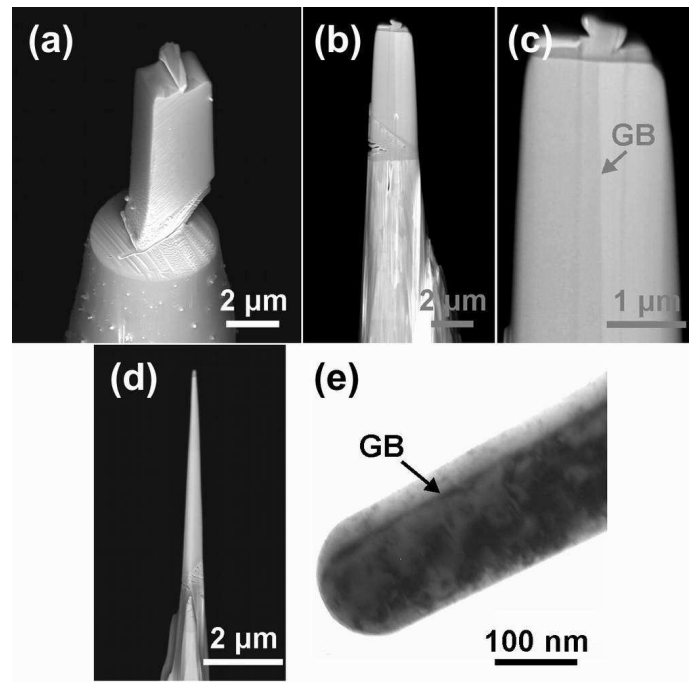

**Figure 4** Shaping of the pillar to form a FIM/TAP tip integrating its W support. (a) Pillar on its W support. (b) Specimen after the annular cuts used to shape the W support. (c) Magnified view of the pillar after polishing cuts revealing the position of the grain boundary. (d) Tip during the final stages of preparation. (e) TEM image of the apex region of the finished tip. The arrow denotes the position of the grain boundary within the sample.

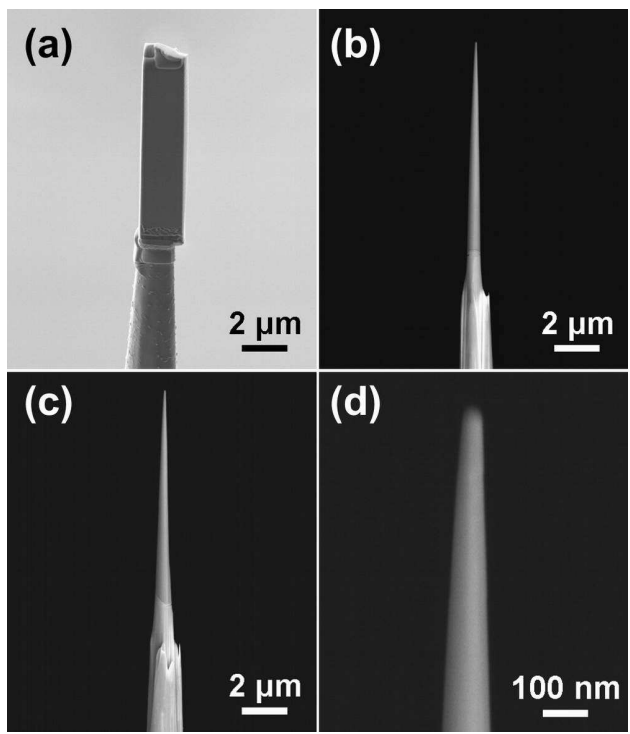

**Figure 5** Shaping of the pillar to form a FIM/TAP tip on a very slender W support. (a) Pillar on its W support. (b), (c) Views of the finished sample showing its cylindrical symmetry. Between panels (b) and (c) the specimen was rotated around its axes by 90°. (d) Magnified image of the apex region. The radius of curvature of the tip apex is well below 50 nm.

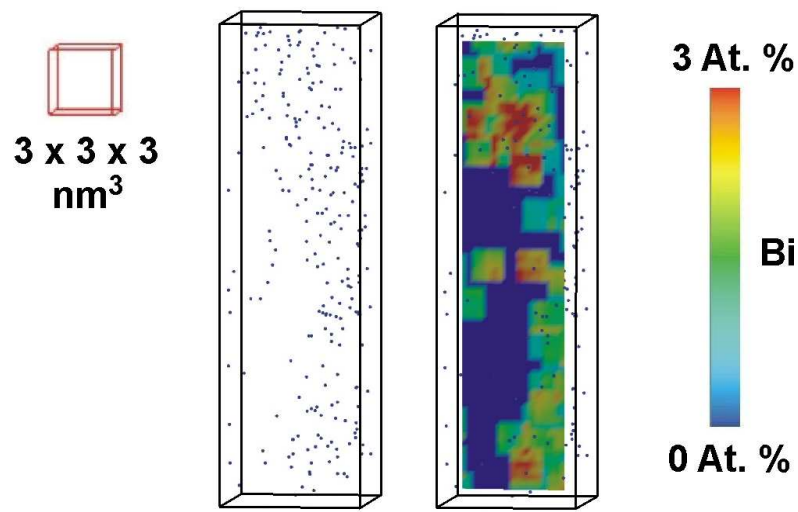

**Figure 6** Three-dimensional TAP reconstruction for a FIB prepared FIM/TAP sample. The region depicted is 25 nm x 7 nm x 7 nm. Each of the blue dots represents an individual Bi atom. For clarity, the Cu atoms are not shown. On the right-hand side a coloured planar concentration profile is presented which reveals more clearly the position of the grain boundary within the analysed volume.